\documentclass[acmtog]{acmart}
\acmSubmissionID{1482}
\usepackage{booktabs} 
\usepackage[ruled]{algorithm2e} 
\usepackage{bm}

\SetAlFnt{\small}
\SetAlCapFnt{\small}
\SetAlCapNameFnt{\small}
\SetAlCapHSkip{0pt}

\acmJournal{TOG}

\begin{document}

\title{Enhancing Reference-based Sketch Colorization via Separating Reference Representations}

\author{Dingkun Yan}
\authornote{Both authors contributed equally to this research.}
\email{
tellurion.kanata@gmail.com}
\affiliation{%
  \institution{Tokyo University of Science}
  \city{Tokyo}
  \country{Japan}
}

\author{Xinrui Wang}
\authornote{Both authors contributed equally to this research.}
\email{secret_wang@weblab.t.u-tokyo.ac.jp}
\affiliation{%
  \institution{The University of Tokyo}
  \city{Tokyo}
  \country{Japan}
}

\author{Zhuoru Li}
\email{hatsuame@gmail.com}
\affiliation{%
  \institution{Project HAT}
  \city{Xiamen}
  \country{China}
  }

\author{Suguru Saito}
\email{suguru@img.cs.titech.ac.jp}
\affiliation{%
  \institution{Tokyo University of Science}
  \city{Tokyo}
  \country{Japan}
}

\author{Yusuke Iwasawa}
\email{iwasawa@weblab.t.u-tokyo.ac.jp}
\affiliation{%
  \institution{The University of Tokyo}
  \city{Tokyo}
  \country{Japan}
}

\author{Yutaka Matsuo}
\email{matsuo@weblab.t.u-tokyo.ac.jp}
\affiliation{%
  \institution{The University of Tokyo}
  \city{Tokyo}
  \country{Japan}
}

\author{Jiaxian Guo}
\email{jiaxianguo07@gmail.com}
\affiliation{%
  \institution{The University of Tokyo}
  \city{Tokyo}
  \country{Japan}
}

\begin{teaserfigure}
        \centering
        \includegraphics[width=0.98\linewidth]{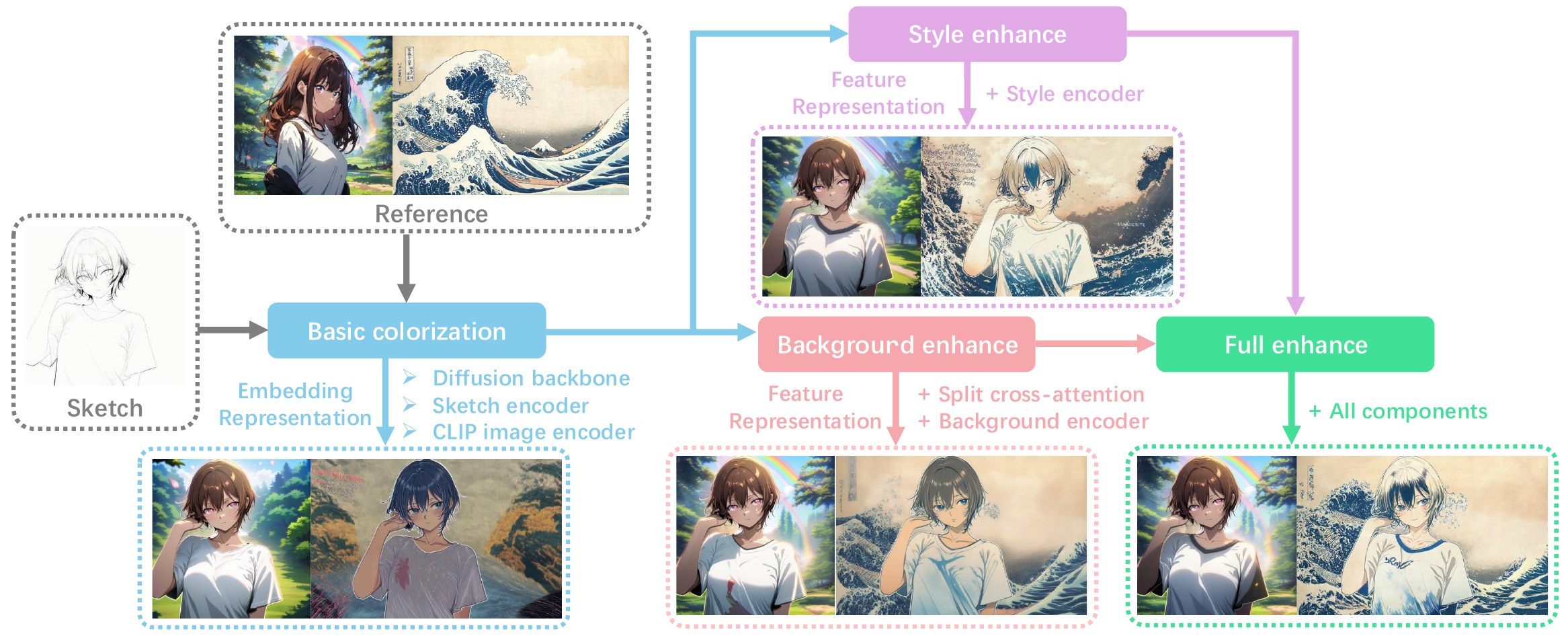}
        \captionof{figure}{We propose a reference-based sketch colorization workflow that employs separate reference representations with respective modules and is trained with a multi-stage schedule. The system achieves state-of-the-art anime-style colorization results without requiring spatial correspondence between inputs.}
        \label{teaserfigure}
\end{teaserfigure}

\begin{abstract}
Reference-based sketch colorization methods have garnered significant attention for the potential application in animation and digital illustration production. However, most existing methods are trained with image triplets of sketch, reference, and ground truth that are semantically and spatially similar, while real-world references and sketches often exhibit substantial misalignment. This mismatch in data distribution between training and inference leads to overfitting, consequently resulting in artifacts and significant quality degradation in colorization results. To address this issue, we conduct an in-depth analysis of the \textbf{reference representations}, defined as the intermedium to transfer information from reference to sketch. Building on this analysis, we introduce a novel framework that leverages distinct reference representations to optimize different aspects of the colorization process. Our approach decomposes colorization into modular stages, allowing region-specific reference injection to enhance visual quality and reference similarity while mitigating spatial artifacts. Specifically, we first train a backbone network guided by high-level semantic embeddings. We then introduce a background encoder and a style encoder, trained in separate stages, to enhance low-level feature transfer and improve reference similarity. This design also enables flexible inference modes suited for a variety of use cases. Extensive qualitative and quantitative evaluations, together with a user study, demonstrate the superior performance of our proposed method compared to existing approaches. Code and pre-trained weight will be made publicly available upon paper acceptance.

\begin{figure*}[t]
    \centering
    \includegraphics[width=\linewidth]{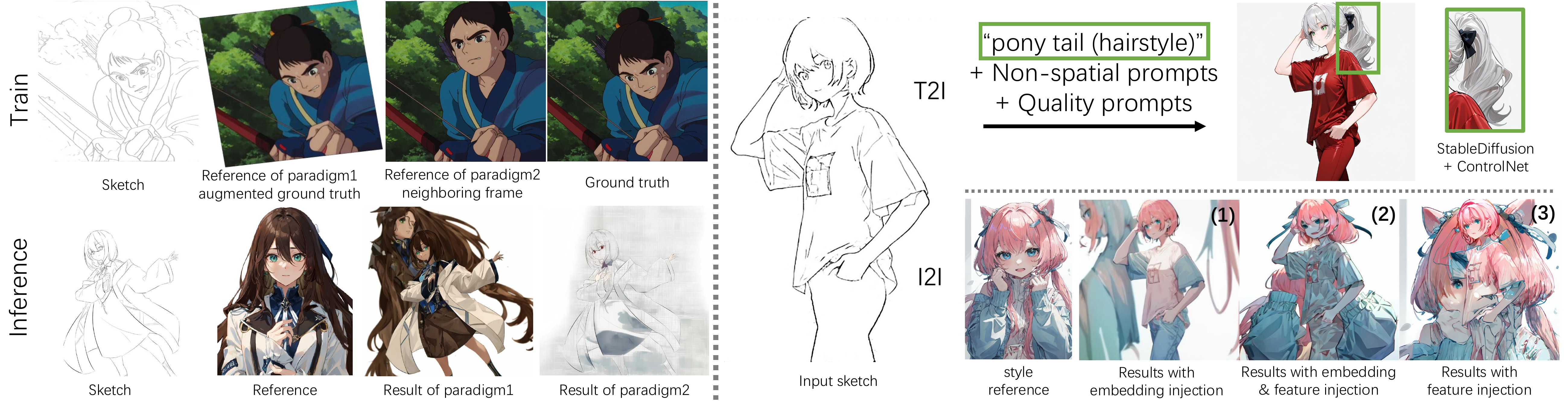}
    \vspace{-2em}
    \caption{Left: data used for training and inference exhibit a significant gap in the semantics and geometry between reference images and sketches due to their different sources. This gap results in severe deterioration in the colorization quality. Right: spatial entanglement caused by this gap when adopting the first training paradigm, which is termed as distribution shift in \cite{cvpr-colorizediffusion}. This issue makes the T2I method mistakenly change the hairstyle in the result, and artifacts in the I2I methods are influenced by the intermedium of reference injection. The artifacts increase when there is more detailed information injected into the common training scheme. Training frames come from movie \textit{Princess Mononoke}}
 \label{entanglement}
\vspace{-1em}
\end{figure*}

\end{abstract}

    
\maketitle

\section{Introduction}
\label{Introduction}


Animation has long been a popular artistic form and in great demand by the audience around the world for decades. Contemporary animation production, however, remains labor-intensive, and increasing market demand is straining studio capacity, posing significant challenges to the industry. Within current animation creation workflows, sketch colorization represents a particularly labor-intensive process, occupying a substantial portion of studio personnel. Consequently, machine learning techniques have been explored to automate this task and alleviate manual effort. 

Early attempts utilizing Generative Adversarial Networks (GANs) \cite{zhang2017style, ZhangLW0L18, zouSA2019sketchcolorization, yan-cgf} yield suboptimal colorization results due to the limitation in their generative capacity. Recently, diffusion models have been applied to sketch colorization because of their ability of synthesizing high-quality images, and methods with images as references are the most popular. Based on the training design, recent methods can be categorized into two types: 1. using reference inputs derived from ground truth \cite{Yan_2025_WACV, ip-adapter, t2i-adapter}; 2. using images with the same identities as reference, such as neighboring frames in a video clip \cite{liu2025manganinja} or manga grids \cite{zhuang2025cobraefficientlineart} and directly transfer low-level representations during training. Shown in Figure \ref{entanglement}, both training paradigms use reference images similar to ground truth as training data, introducing distribution shift during inference, as reference images and sketches are usually spatially and semantically less relevant for inference. The first kind of methods usually experience spatial artifacts called spatial entanglement \cite{Yan_2025_WACV}, showing as additional characters or body parts in the background region or unexpected changes of sketch semantics, while the second kind of methods tend to synthesize insufficiently colorized results with blurry details.

This paper proposes a workflow that follows the first training paradigm and aims to eliminate spatial entanglements and enhance colorization qualities. We start with an analysis of the distribution shift and its main artifacts. Shown in Figure \ref{entanglement}, the discrepancy between the training and inference data distributions leads to artifacts and severe deterioration in image quality for inputs with misaligned semantics or structures. 
This occurs because the intermediums used to inject reference information into diffusion models during training, termed \textbf{\textit{reference representations}} in this paper, contain not only color and style information used for colorization, but also spatial content that might contradicts with the sketch semantics. When injected into the diffusion backbone, the misaligned content information causes incorrectly colorized regions beyond the control of the sketches. We further discovered that different reference representations used for reference injection are causing corresponding distinct artifacts.
Embeddings with higher-level semantic information as reference representations tend to produce results with fewer artifacts but blurry textures, while using latent features with richer details and lower-level semantics as reference representations yields results with higher similarity but more pronounced artifacts. Denoted ``spatial entanglement,'' these artifacts are illustrated with corresponding reference representations in the right half of Figure \ref{entanglement}.

\begin{figure*}[htb]
    \centering
    \includegraphics[width=1\linewidth]{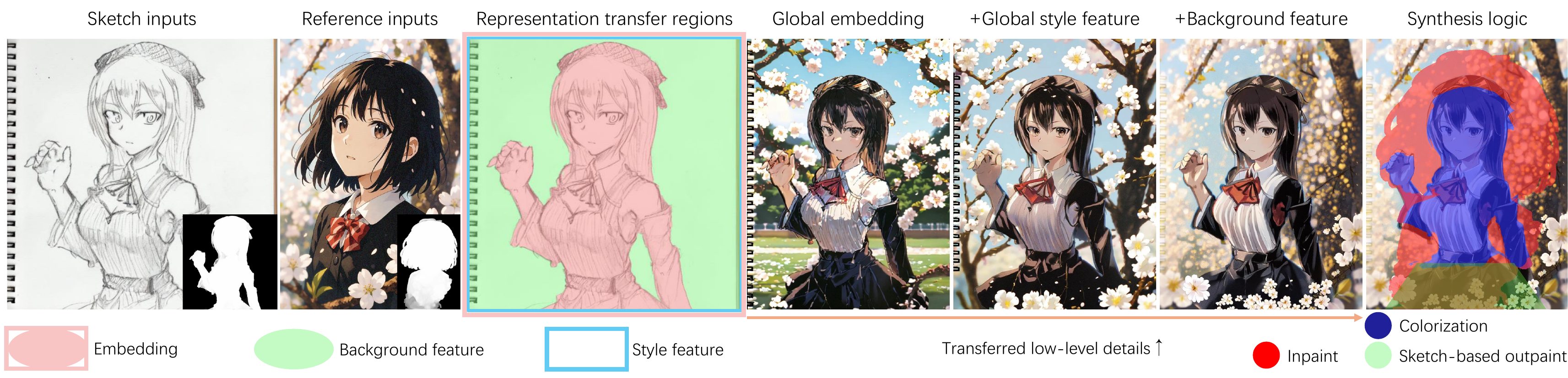}
    \vspace{-1.5em}
    \caption{Illustration of the proposed reference-based sketch colorization workflow. To eliminate artifacts and enhance colorization quality, we separate colorization into three parts, leveraging foreground masks extracted from the reference and sketch inputs: embedding guidance for sketch-covered regions, style modification for global details, and low-level transfer for non-sketch regions. Moreover, the network should be able to properly inpaint the missing regions based on neighboring features in the sketch and reference images. As highlighted by red, the proposed network inpaints the skirt based on prior knowledge from the sketch and the flowers based on neighboring features from the reference.}
    \label{colorization-logit}
    \vspace{-0.5em}
\end{figure*}

To eliminate spatial entanglement artifacts, we start with employing a split cross-attention mechanism \cite{cvpr-colorizediffusion} with embedding reference representations. Spatial masks segment the foreground and background regions in both the sketch and reference images, allowing the network to process their semantics independently and avoid mutual interference. However, this mechanism degrades the texture and details of colorized backgrounds. Therefore, we introduce a background encoder collaborated with feature reference representations to facilitate the transfer of fine-grained details from the reference image, especially the background regions. Furthermore, a style encoder is introduced to better transfer style information such as tone and textures to the full image. To preserve the network's understanding of the sketch semantics and prevent different modules from influencing each other, we design a multi-stage training strategy that trains each component in different stages.
Additionally, we propose two novel training techniques—character-mask merging and background bleaching—which are essential for effective background-specific feature injection. Further details are provided in the supplementary materials.

We conduct extensive experiments, including qualitative evaluations showing that our method generates high-quality results and faithfully transfers the color and textures from reference images while avoiding spatial entanglement, quantitative comparisons against existing methods validating the superiority of the proposed approach, and a comprehensive ablation study demonstrating the contribution of each component in mitigating spatial entanglement and artifacts in various scenarios. Finally, user studies shows that users prefer our method over existing methods.

In summary, our contributions are threefold: (1) We provide a detailed analysis of reference-based sketch image colorization methods and identify the underlying causes and manifestations of artifacts. (2) We propose a novel colorization framework with components designed to address specific artifact types, resulting in effective mitigation and high-quality colorization without requiring well-aligned input pairs. (3) Extensive experiments demonstrate the superiority of our method over existing approaches through qualitative and quantitative comparisons, as well as a perceptive user study.

\section{Related Work}
\subsection{Latent Diffusion Models}

Diffusion Probabilistic Models \cite{HoJA20,0011SKKEP21} are a class of latent variable models inspired by nonequilibrium thermodynamics \cite{Sohl-DicksteinW15} and have achieved great success in image synthesis and editing. Compared to Generative Adversarial Networks (GANs)\cite{GoodfellowPMXWOCB14,KarrasLA19,KarrasLAHLA20}, Diffusion Models excel at generating high-quality images with various contexts and stronger control over different conditional guidance. However, the autoregressive denoising process of diffusion models, typically computed with a U-Net \cite{RonnebergerFB15} or a Diffusion Transformer (DiT) \cite{DiT,pixart}, incurs substantial computational costs. To reduce this cost, Rombach et al. proposed Stable Diffusion (SD) \cite{RombachBLEO22,sdxl}, a class of Latent Diffusion Models (LDMs) that performs a denoising process in a perceptually compressed latent space with a pair of pre-trained Variational Autoencoder (VAE). Also, studies on accelerating the denoising process have demonstrated effectiveness \cite{SongME21,0011SKKEP21,0011ZB0L022,abs-2211-01095}. In this paper, we adopt SD as the backbone, utilize the DPM++ solver \cite{abs-2211-01095,0011SKKEP21,KarrasAAL22} as the default sampler, and employ classifier-free guidance \cite{DhariwalN21,abs-2207-12598} to strengthen the transfer performance.

\begin{figure*}[t]
    \centering
    \includegraphics[width=1\linewidth]{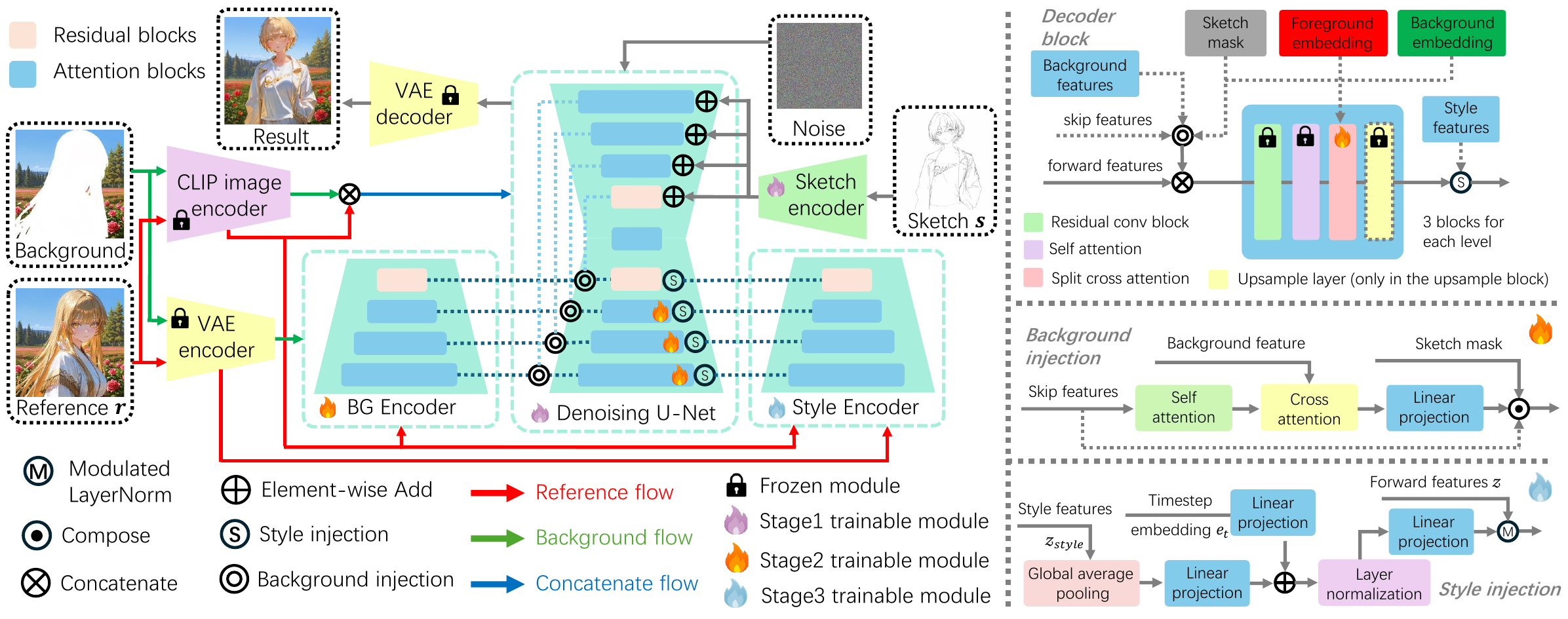}
    \vspace{-2em}
    \caption{Illustration of the proposed framework. Left shows the whole framework. The CLIP image encoder and the VAE encoder are fixed during training. The extracted image embeddings and latent images are injected into the corresponding modules in the same way as standard LDM. The denoising U-Net, the background encoder and the style encoder are trained separately in 3 stages. Right shows the detailed architectures of the decoder block, style injection block, and background injection block. The calculation of ``compose'' is implemented as Eq. \ref{warp-module}. Detailed pipelines of stage 1 and stage 2 are included in the supplementary materials.}
    \label{framework}
    \vspace{-0.5em}
\end{figure*}

\subsection{Sketch colorization}

Sketch colorization has been an active research area in computer vision. Early approaches started from interactive optimization \cite{SykoraDC09}, deep learning then emerged as the dominant paradigm, enabling the synthesis of high-quality, high-resolution color images \cite{ZhangLW0L18, KimJPY19, li2022eliminating, yan2024colorizediffusion}. Based on the guiding modality, existing methods can be broadly categorized into three groups: text-prompted \cite{KimJPY19,yan-cgf,controlnet-iccv}, user-guided \cite{ZhangLW0L18,s2pv5}, and reference-based \cite{li2022eliminating,yan-cgf,animediffusion}. User-guided methods offer fine-grained control through direct user input (e.g., spots, sprays), but their reliance on manual intervention hinders their applicability in automated pipelines. Text-prompted methods, driven by advancements in text-to-image diffusion models, have gained significant traction, yet it is challenging to precisely control colors, textures, and styles using text prompts. 

Image-referenced methods have also benefited from advancements in diffusion models and image control techniques \cite{controlnet-iccv,controllllite,controlnet-v11,t2i-adapter,ip-adapter}. However, a critical challenge remains: effectively addressing the spatial and semantic conflicts between diverse reference images with the often sparse and abstract nature of sketches during inference. Existing methods typically require extracted sketches with highly-matched references as input pairs to achieve satisfying results \cite{li2022eliminating,yan-cgf,animediffusion,tooncraft,meng2024anidoc,lvcd}, limiting their generalizability and potential applications. While ColorizeDiffusion \cite{yan2024colorizediffusion} demonstrated significant progress in colorization quality, it still grapples with spatial entanglement, as visualized in Figure \ref{entanglement}, and struggles to accurately transfer fine-grained details. In this paper, we address these limitations by introducing a novel step-by-step training strategy within a refined colorization framework designed to explicitly mitigate spatial entanglement.

\section{Method}
To eliminate the spatial entanglement caused by distribution shift and enhance the generalization capabilities and performance of reference-based sketch colorization, we propose a multi-stage framework that employs both embedding representations that contain high-level semantic information and feature representation that contain low-level texture information, and colorize sketches in a coarse-to-fine manner. In the first stage, we train the diffusion backbone with the embedding representation. In the second stage, the backbone is fixed and a background encoder is trained with feature representation to synthesize the finer details in the background region, and split cross-attention mechanism \cite{cvpr-colorizediffusion} is exploited with spatial masks segmenting foreground and background of reference images and sketches to separately inject reference information of corresponding regions into the diffusion backbone. Especially, we design a character-mask merging strategy together with a background bleaching strategy to inpaint the possible overlap regions of reference masks and sketch masks. In the third stage, we freeze all other modules and optimize an additional style encoder with feature representation to reduce content transfer and further enhance global style textures. The colorization workflow and training pipeline are visualized in Figure \ref{colorization-logit} and Figure \ref{framework}, respectively. 

The framework consists of a diffusion backbone, a variational autoencoder (VAE), a sketch encoder, a CLIP image encoder, a style encoder, and a background encoder. It leverages a sketch image $\bm{X_s} \in \mathbb{R}^{w_s \times h_s \times 1}$, a reference image $\bm{X_r} \in \mathbb{R}^{w_r \times h_r \times c}$, a sketch mask $\bm{X_{sm}} \in \mathbb{R}^{w_s \times h_s \times 1}$ and a reference mask $\bm{X_{rm}} \in \mathbb{R}^{w_r \times h_r \times 1}$ as inputs, and returns the colorized result $\bm{Y} \in \mathbb{R}^{w_s \times h_s \times c}$, with $w$, $h$ and $c$ representing the width, height and channel of the images.

\begin{figure*}[t]
    \centering
    \includegraphics[width=1\linewidth]{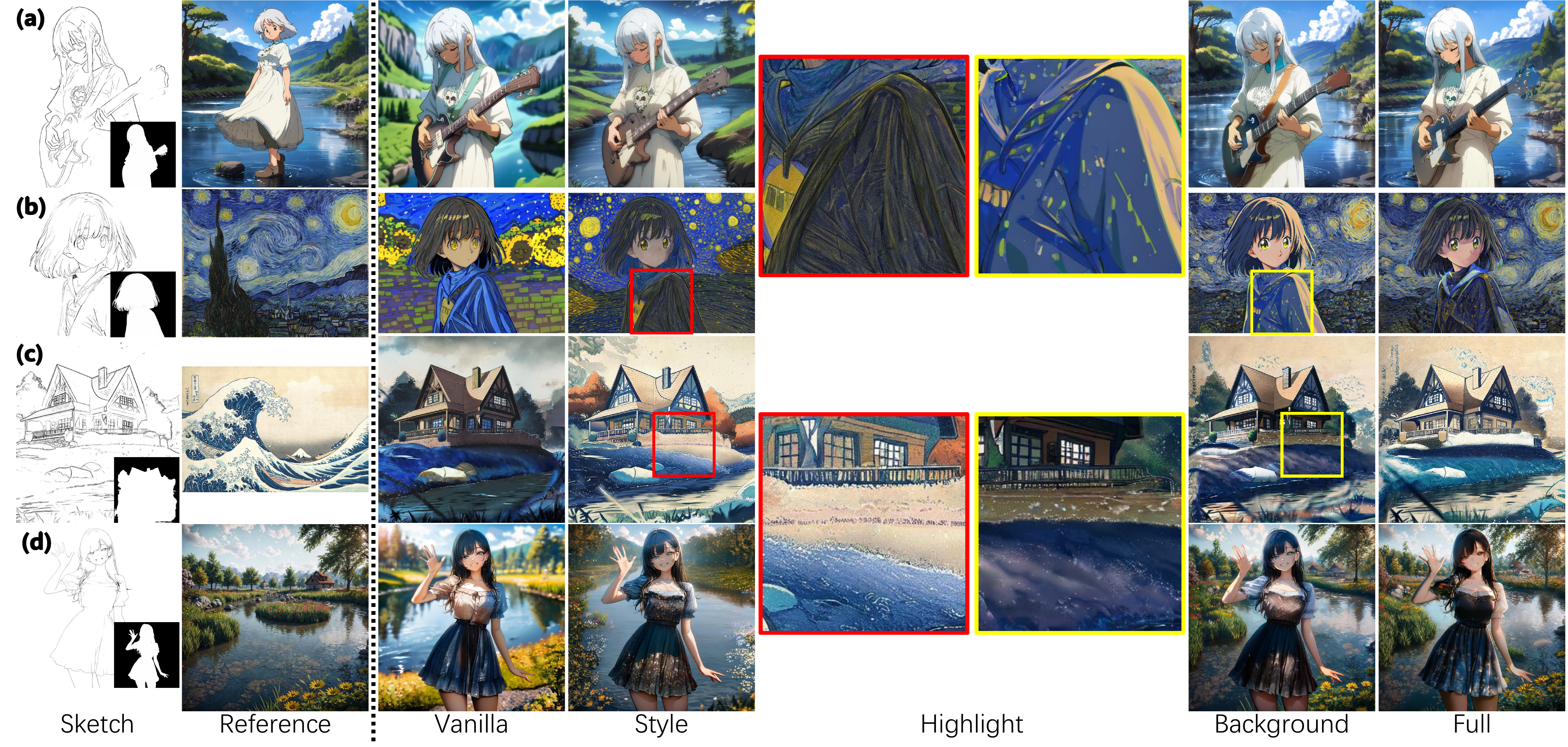}
    \vspace{-2em}
    \caption{Colorization results with different encoders. \textit{Style enhance} modes transfer low-level features globally, while \textit{background enhance} modes transfers to backgrounds. We highlight the differences of stroke details inside foregrounds between \textit{style} and \textit{background} modes from rows (b) and (c). High-resolution images and failure cases are available in the supplementary materials.}
    \label{inference-mode}
\end{figure*}

\subsection{Extraction of reference representations}

Existing sketch colorization methods typically employ feature extraction networks to derive reference intermediums from reference images and inject them into a diffusion backbone. Illustrated in Figure \ref{entanglement}, we utilize both image embeddings \cite{ip-adapter, yan2024colorizediffusion} and latent features \cite{lvcd, tooncraft} as reference representations in the proposed method.

We exploit the OpenCLIP-H image encoder\cite{RadfordKHRGASAM21,openclip,openclip-2,schuhmann2022laionb} to extract image embeddings as the embedding representations. The pre-trained ViT-based image encoder network extracts 2 kinds of image embeddings: the CLS embeddings $\bm{E_{cls}} \in \mathbb{R}^{bs \times 1 \times 1024}$ and the local embeddings $\bm{E_{local}} \in \mathbb{R}^{bs \times 256 \times 1024}$, where $bs$ represent the batch size. Previous image-guided diffusion models \cite{ip-adapter, instantstyle} commonly utilize CLS embeddings as color or style references, which are connected to text-level notions and projected to CLIP embedding space for image-text contrastive learning, with spatial information compressed as spatial semantics. ColorizeDiffusion \cite{yan2024colorizediffusion}, on the other hand, reveals that local embeddings also express text-level semantics and, meanwhile, express more spatial details regarding textures, strokes, and styles, enabling the network to generate better reference-based results with finer details. Consequently, the proposed method follows \cite{yan2024colorizediffusion} to adopt local embeddings as a reference representation. The extracted embedding representations are then injected into the diffusion U-Net through split cross-attention.

Nevertheless, the inherent lack of low-level, detailed reference information in image embeddings can lead to blurry textures in the colorized output. For sketches depicting isolated figures or featuring simple background lines, the information encoded within image embeddings may be insufficient to reconstruct a meaningful and visually compelling background. To address this limitation, we incorporate a style encoder and a background encoder to extract latent features as a reference representation to facilitate the reference information injection. The background encoder and the style encoder are initialized from the encoder of diffusion U-Net trained in stage 1 and optimized in stage 2 and stage 3, respectively, and the features extracted by them are injected into the diffusion backbone through corresponding injection modules as reference representations. 


\subsection{Feature injection modules}
\label{background_encoder}

In existing colorization methods, spatial entanglements are mainly represented as foreground exceeding boundaries and appearing in the background, such as extra characters or body parts. An efficient way to eliminate the phenomena is to separately process the foreground and background with a split cross-attention mechanism \cite{cvpr-colorizediffusion}, where sketch-guided regions are regarded as foreground and all other regions are regarded as background. However, utilizing this mechanism usually degrades the quality and textures of colorization results. To further enhance textures and semantics synthesis, particularly for sketches with sparse or absent background lines, we introduce a background encoder and a style encoder to transfer features with detailed information to backgrounds and facilitate the synthesis of fine textures. 

As illustrated in Figure \ref{framework}, we design a background injection module and a style injection module to integrate the output of background encoder and style encoder respectively into the denoising U-Net decoder at corresponding levels. We denote the sketch mask as $\bm{m_{s}}$, the user-defined foreground threshold for the sketch input as $ts_{s}$, computation of the transformer block as $\mathcal{W}(\cdot)$, the skip feature from the encoder of the denoising U-Net as $\bm{z}_{skip}$, and the feature from the background encoder as $\bm{z}_{bg}$, the calculation of the background injection module is formulated as:
\begin{equation}
    \bm{z}_{skip} = \begin{cases}
    \bm{z}_{skip} & \text{if $\bm{m_{s}} > ts_{s}$}\\
    \mathcal{W}(\bm{z}_{skip},\bm{z}_{bg}) & \text{if $\bm{m_{s}}\leq ts_{s}$}.
    \end{cases}
    \label{warp-module}
\end{equation}
For the style injection module, we adopt adaptive normalization \cite{HuangB17} to enhance global style transfer. Given the forward feature in denoising U-Net as $\bm{z}$, the style feature from the style encoder as $\bm{z}_{style}$, global average pooling as $\text{GAP}(\cdot)$, timestep embedding as $\bm{e}_{t}$, the style modulation $\mathcal{M}(\cdot)$ can be formulated as
\begin{equation}
    \mathcal{M}(\bm{z},\hat{\bm{z}}_{scale},\hat{\bm{z}}_{shift},\bm{e}_{t})=\bm{z}\cdot(1+\hat{\bm{z}}_{scale})+\hat{\bm{z}}_{shift},
    \label{style_modulation-eq1}
\end{equation}
where $\hat{\bm{z}}_{scale}$ and $\hat{\bm{z}}_{shift}$ are obtained via linear projections from $\bm{z}_{style}$, conditioned on the timestep embedding $\bm{e}_{t}$.

\begin{figure}[t]
    \centering
    \includegraphics[width=\linewidth]{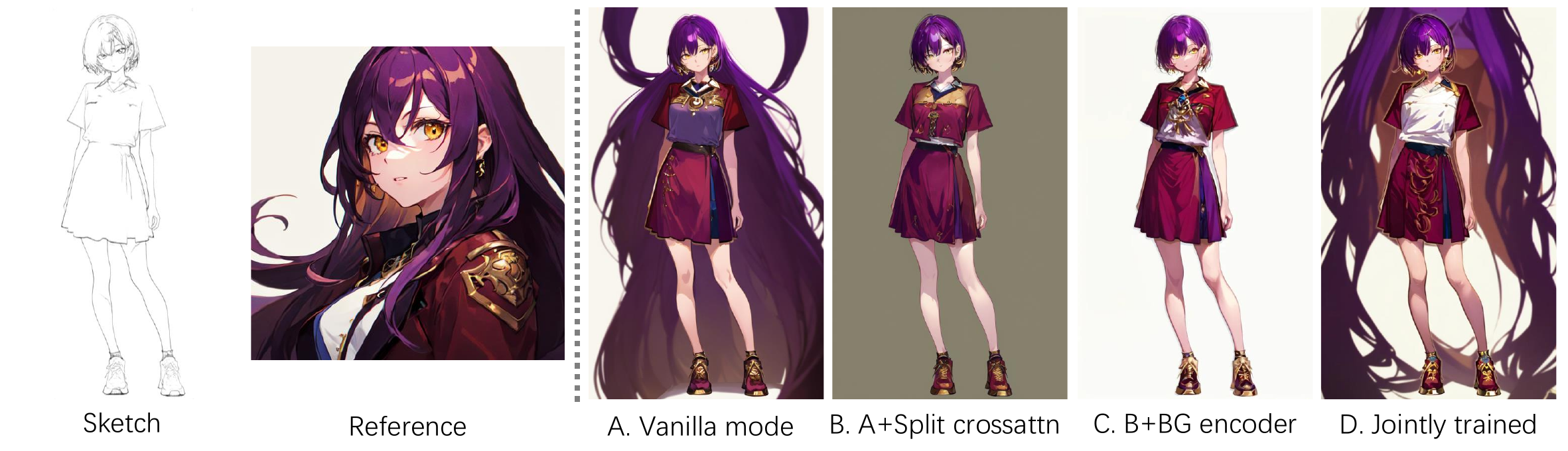}
    \vspace{-1.5em}
    \caption{The separate processing effectively eliminates spatial entanglement artifacts in background regions, while the background encoder further enhances background synthesis. However, jointly training these components with the denoising U-Net negates these improvements by preventing the necessary disentanglement.}
    \label{ablation}
    \vspace{-1.5em}
\end{figure}

\subsection{Multi-step training strategies}
\label{style_encoder}
The proposed framework consists of multiple components, with each component playing a different role and having unique effects on the final results. If the framework is jointly trained with the same ground truth, these components are likely to influence each other and fail to learn the information as designed. 

To tackle this issue, we further design a multi-stage training strategy to train the diffusion backbone, the background encoder, and the style encoder separately: 1. Colorization pre-training stage: we only optimize the sketch encoder and denoising U-Net in this stage. Following the experience of \cite{yan2024colorizediffusion}, we train the network with a two-stage noisy training-refinement scheduler. A dynamic reference drop of 80\% for the noisy training stage and 50\% for the refinement stage is adopted to avoid severe deterioration in the segmentation and perceptual quality of results; 2. Foreground-background separated training stage: we add the split cross-attention module and the background encoder into the framework and optimize them with other parameters frozen. This stage helps eliminate spatial entanglement caused by the reference embeddings and enhances the synthesis of backgrounds; 3. Hybrid training stage of style encoder: the style encoder is optimized with the background encoder and split cross-attention not trained but randomly activated at a rate of 50\% to generate extra conditions for the denoising backbone and other parameters frozen. In stage 2 and stage 3, the reference embeddings for denoising U-Net are dropped at a fixed rate of 50\%. Given noise $\epsilon$, sketch $\bm{s}$, ground truth $\bm{y}$, encoded latent representations (forward features) $\bm{z}_{t}$ at timestep $t$, VAE encoder $\mathcal{E}$, denoising U-Net and sketch encoder $\theta$, background encoder with background injection $\varphi_{bg}$, style encoder with style injection $\varphi_{style}$, and CLIP image encoder $\phi$, the training objective for all training stages can be defined as 
\begin{equation}
    \mathcal{L}(\vartheta)=\mathbb{E}_{\mathcal{E}(\bm{y}),\epsilon,t,\bm{s},\bm{c}}[\|\epsilon-\epsilon_{\theta}(\bm{z}_{t},t,\bm{s},\bm{c})\|^{2}_{2}],
    \label{diffusion-loss}
\end{equation}
where $\vartheta$ and $\bm{c}$ represent the optimization targets and conditional inputs, and a detailed explanation for each stage is as follows. \textbf{Stage 1}: $\vartheta$ represents the denoising U-Net and the sketch encoder, and $\bm{c}$ represents image embeddings $\bm{e}=\phi(\bm{r})$; \textbf{Stage 2}: $\vartheta$ represents the background encoder and injection modules, as well as LoRAs inside split cross-attention layers, and $\bm{c}$ represents background embeddings $\bm{e}_{bg}$, sketch mask $\bm{m}_{s}$, and background features $\bm{z}_{bg}=\varphi_{bg}\big(\mathcal{E}(\bm{r}_{bg}),\bm{e}\big)$; \textbf{Stage 3}: $\vartheta$ represents the style encoder and injection modules, and $\bm{c}$ represents style features $z_{style}=\varphi_{style}\big(\mathcal{E}(\bm{r}),\bm{e}\big)$. As ground truths are directly used as references, $\bm{r}$ can be replaced with $\bm{y}$ throughout all equations.


Specifically, the style encoder is trained in Stage 3, following the training of the background encoder. This scheduling helps prevent the style encoder from capturing excessive content information.

\section{Experiment}

\begin{figure}[t]
    \centering
    \includegraphics[width=\linewidth]{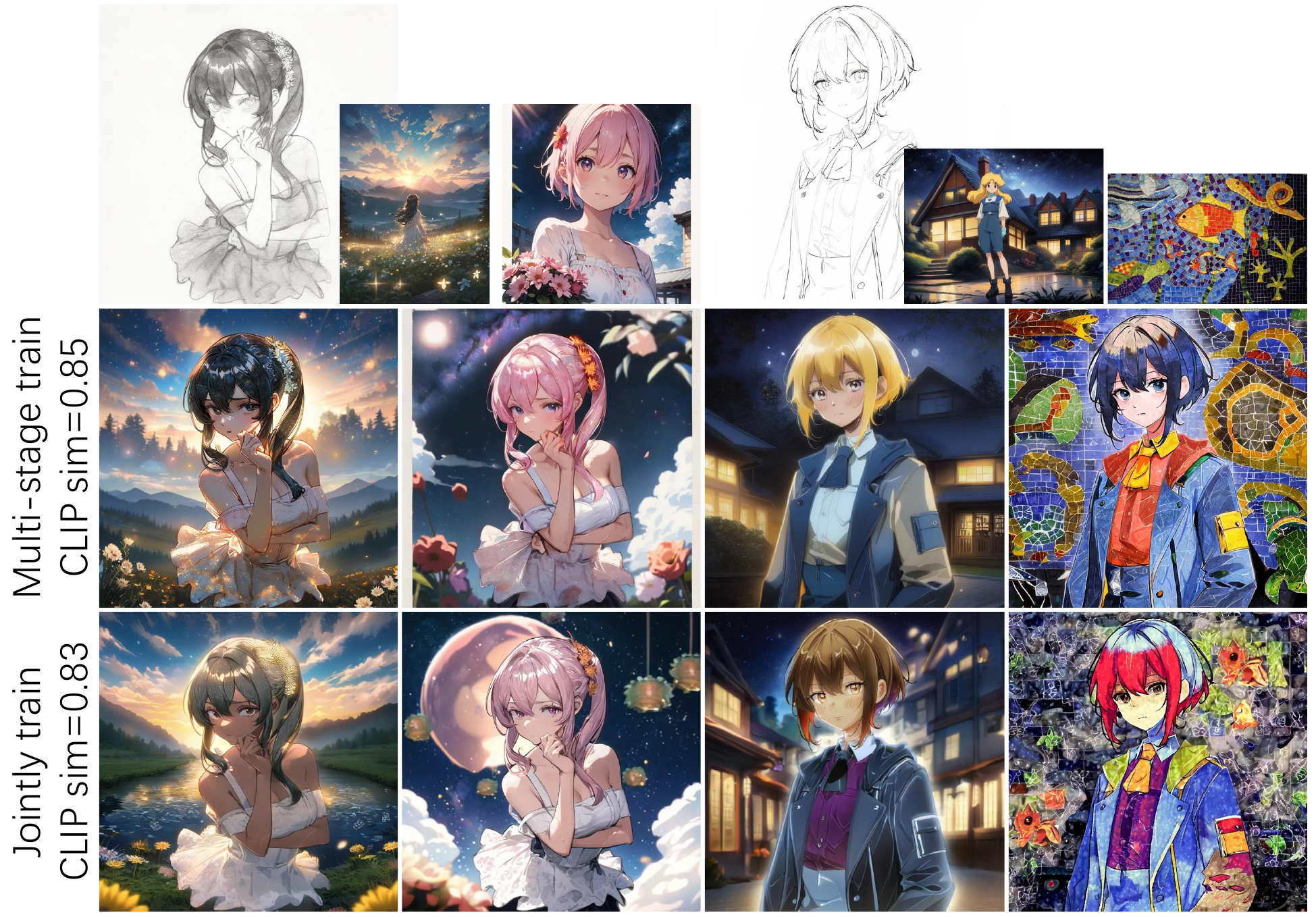}
    \vspace{-1.5em}
    \caption{Ablation study regarding training strategy for the style injection. We calculated 5K CLIP similarity to evaluate their transfer performance quantitatively in this comparison, and all results were generated using \textit{style enhance} mode.}
    \label{strategy-ablation}
    \vspace{-1.5em}
\end{figure}

\begin{figure*}[t]
    \centering
    \includegraphics[width=1\linewidth]{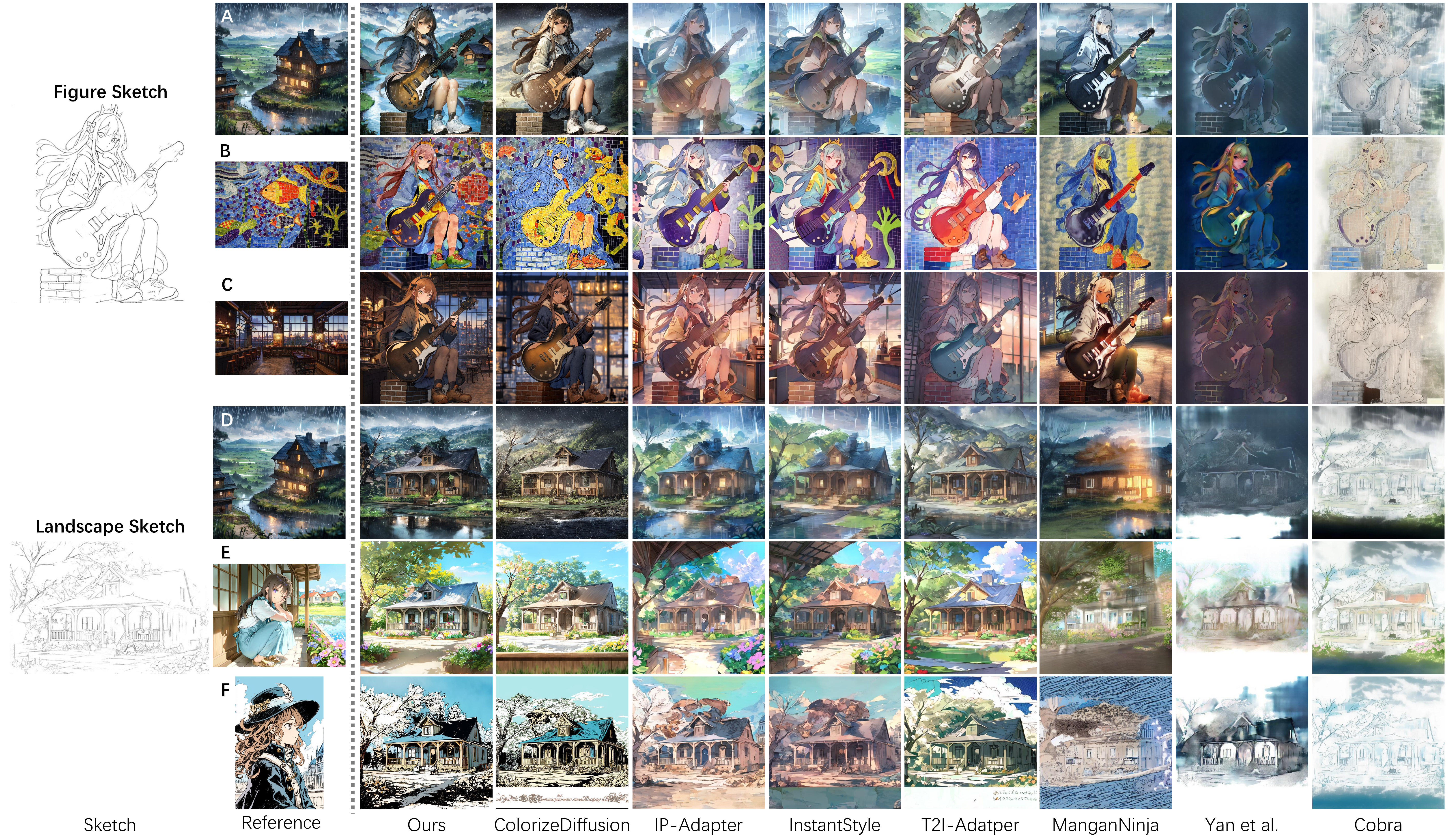}
    \vspace{-2em}
    \caption{Cross-content transfer results. Raw A-C show results of figure sketches, where ours significantly outperform in segmentation and colorization. Raw D-F show results of landscape sketches, where ours demonstrate better composition and transfer performance, as well as outpainting for non-sketch regions.}
    \label{cross-qualitative}
    \vspace{-1em}
\end{figure*}

\subsection{Implementation}

\noindent\textbf{Dataset, configurations, and environment.} 
We curated a dataset of over 6M (sketch, color, mask) image triples from Danbooru \cite{danbooru2021}, encompassing a diverse range of anime styles for illustration-format images. Sketches were generated by jointly using \cite{sketchKeras} and \cite{xiang2022adversarial}, while masks were produced using \cite{anime-segmentation}. The dataset was divided into a training set and a validation set of 54,000 triples without overlap. The training was conducted on 8x H100 (94GB) GPUs utilizing Deepspeed ZeRO2 \cite{deepspeed} and the AdamW optimizer \cite{KingmaB14, LoshchilovH19} with a total batch size of 256, a learning rate of 0.00001, and betas of (0.9, 0.999). All training inputs were initially resized to 800$\times$800 and then randomly cropped to 768$\times$768, with the exception of reference images, which were resized directly. More details are described in the supplementary material.\\

\noindent\textbf{Inference mode.} To address a variety of use cases, the proposed framework provides three alternative inference schemes by activating different components: the \textbf{\textit{Vanilla}} mode, the \textbf{\textit{Style enhance}} mode, and the \textbf{\textit{Background enhance}} mode. \textbf{\textit{Vanilla}} mode only activates diffusion backbone and sketch encoder, and discards split cross-attention and additional encoders. It is thus vulnerable to spatial entanglement and recommended for sketches with complicated compositions, such as landscapes.
\textbf{\textit{Style enhance}} mode additionally activates the style encoder based on the \textit{Vanilla} mode, and is able to transfer fine-grained textures and strokes.
\textbf{\textit{Background enhance}} mode additionally activates the background encoder and the split cross-attention layers based on the \textit{Vanilla} mode. As the foreground and background are explicitly separated by spatial masks, it effectively eliminates spatial entanglement and generates backgrounds with fine details and textures.
The two enhancement modes can be jointly activated as \textbf{\textit{full enhance}} to achieve the best transfer performance. Owing to the generalization ability of the training scheme, the proposed framework can be applied to various sketch-based style transfer applications leveraging sketch extractors. Detailed discussion with examples are given in the supplementary materials.



\begin{figure*}[t]
    \centering
    \includegraphics[width=1\linewidth]{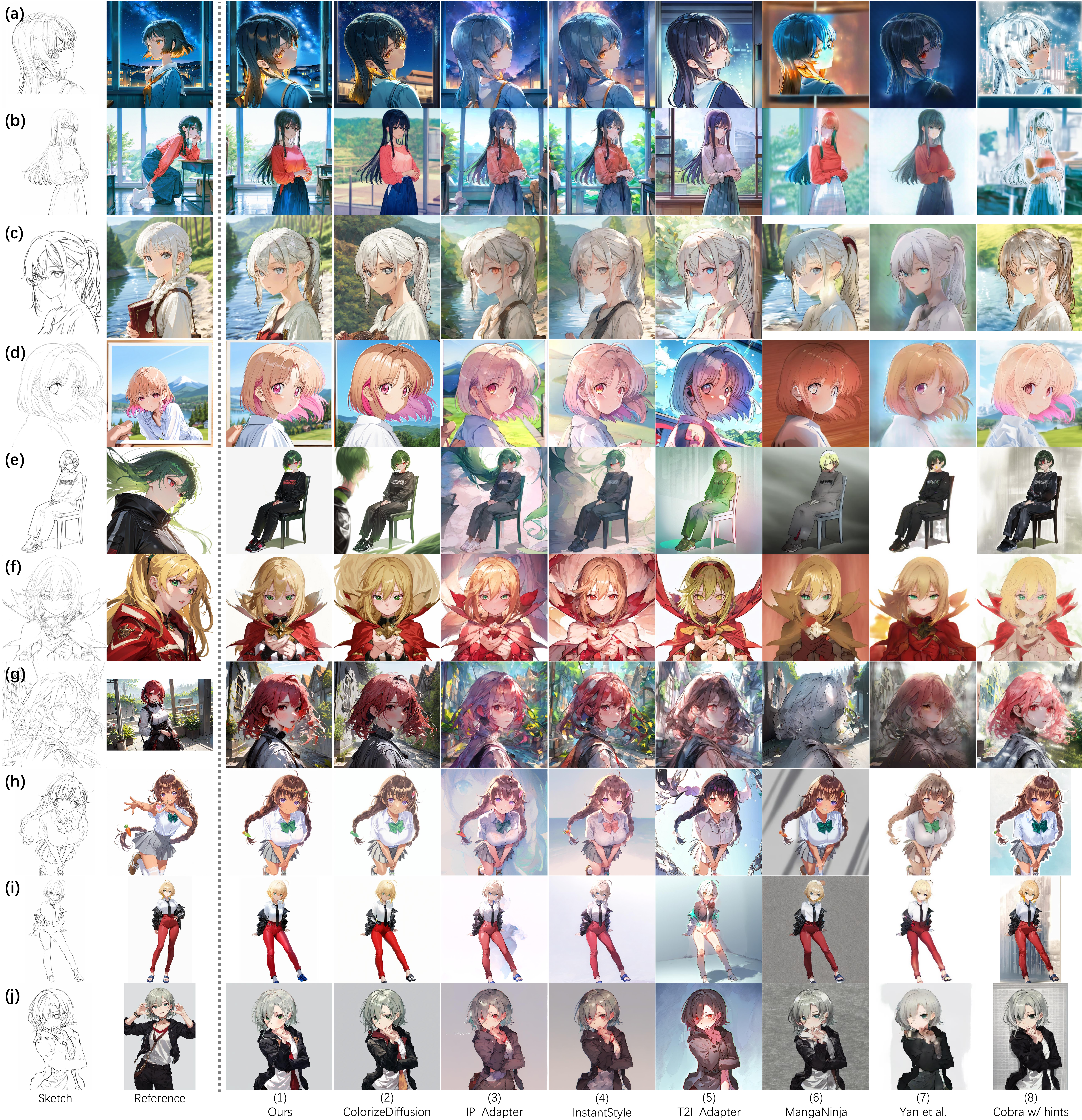}
    \vspace{-1em}
    \caption{Qualitative comparisons regarding figure colorization. Different from recent colorization baselines \cite{liu2025manganinja,zhuang2025cobraefficientlineart,cvpr-colorizediffusion}, the proposed method is demonstrated to be superior in the quality and similarity of colorization without having spatial entanglement and requiring inputs to have semantically or spatially similarity. High-resolution images and user hints used for Cobra \cite{zhuang2025cobraefficientlineart} are contained in the supplementary materials.}
    \label{qualitative}
    \vspace{-1em}
\end{figure*}

\subsection{Ablation study}
\noindent\textbf{Artifacts removal.} We first validate the effectiveness of the proposed framework and training strategy in eliminating background artifacts caused by spatial entanglement \cite{Yan_2025_WACV,cvpr-colorizediffusion}. We set up four ablation frameworks: 1) \textit{vanilla} mode, 2) \textit{vanilla} mode with split cross-attention and trainable LoRAs, 3) \textit{background enhance} mode, and 4) ablation result if related components are jointly optimized in stage 1. 

We show a qualitative comparison in Figure \ref{ablation} to validate the effectiveness of the proposed modules. The ablation frameworks generated additional hair in (a) and (d).\\

\noindent\textbf{Inference modes.} Figure \ref{inference-mode} illustrates the distinctions between inference modes. In \textit{vanilla} mode, colorization relies solely on local image embeddings from the CLIP image encoder. This mode resembles T2I generation, primarily retrieving color attributes from the training data rather than transferring them from reference images. Consequently, \textit{vanilla} mode struggles to replicate intricate stroke details from references that deviate from the anime style, as observed in rows (b)-(d). In contrast, \textit{style enhance} and \textit{background enhance} modes utilize feature representations, enabling the framework to transfer low-level details from references for global style and background, respectively.\\

\noindent\textbf{Training strategy for style encoder.} Aside from diminishing background artifacts and enabling different inference modes for various use cases, the multi-stage training approach is crucial for disentangling style modulation from colorization optimization. To underscore the significance of the multi-stage strategy, we conducted an ablation study. A baseline model was trained for an equivalent GPU time as our full-stage training, but with the style encoder jointly optimized alongside the sketch encoder and the denoising U-Net in the first stage, rather than separately in the final stage. A comparison is presented in Figure \ref{strategy-ablation}. Such joint optimization is problematic because style modulation can hinder the optimization of embedding transfer, as style features stem from low-level visual information and inadvertently encode identity and color semantics. This increased susceptibility typically results in the jointly-trained model performing worse than the model trained via our multi-stage approach during inference, since low-level features are more likely to overfit to training data.

\begin{table*}[t]
    \centering
    \caption{Quantitative comparison on $768^{2}$ resolution between the proposed model and baseline methods. \dag: These evaluations randomly selected color images as references, making them close to real-application scenarios. \ddag: Ground truth color images were deformed to obtain semantically paired and spatially similar references for evaluations. \S: Tested at $512^{2}$ resolution.}
    \vspace{-0.5em}
    \begin{tabular}{|c|c|c|c|c|c|c|c|c|c|}
        \hline
        \multicolumn{5}{|c|}{Method} & {\dag Aesthetic $\uparrow$} & {\dag FID $\downarrow$} & {\ddag PSNR$\uparrow$} & {\ddag MS-SSIM$\uparrow$} & {\ddag CLIP similarity$\uparrow$}\\
        \hline
        \multicolumn{5}{|c|}{Ours} & \textbf{5.1859} &\textbf{5.6330} & 29.3626 & \textbf{0.7081} & \textbf{0.9056} \\
        \hline
        \multicolumn{5}{|c|}{\textit{ColorizeDiffusion}} & 4.8351 & 9.6423 & 28.7215 & 0.5899 & 0.8753 \\
        \hline
	\multicolumn{5}{|c|}{\textit{IP-Adapter}} & 4.6627 & 38.9232 & 28.5124 & 0.5464 & 0.8632 \\
        \hline
        \multicolumn{5}{|c|}{\textit{InstantStyle}} & 4.7150 & 40.2134 & 28.0921 & 0.4467 & 0.8039 \\
	\hline
        \multicolumn{5}{|c|}{\textit{T2I-Adapter}} & 4.2647 &41.1569 & 28.1321 & 0.3194 & 0.7134 \\
        \hline
        \multicolumn{5}{|c|}{\S\textit{MangaNinja}} & 4.1738 & 42.9741 & \textbf{29.5741} & 0.6715 & 0.7304 \\
	\hline
        \multicolumn{5}{|c|}{\S Yan et al.} & 4.7923 &27.0032 & 29.1293 & 0.5239 & 0.8894\\
	\hline
	\end{tabular}
    \label{quantitative}
    \vspace{-0.5em}
\end{table*}

\subsection{Comparison with baseline}
We compare our method with existing reference-based sketch image colorization methods \cite{yan-cgf,liu2025manganinja,cvpr-colorizediffusion,ip-adapter,controlnet-iccv,t2i-adapter,zhuang2025cobraefficientlineart} to demonstrate the superiority of the proposed framework. Depending on training schemes introduced in Section \ref{Introduction}, baseline image-guided methods can be categorized into two classes: 1. training with augmented ground truth as references \cite{cvpr-colorizediffusion,ip-adapter,t2i-adapter,yan-cgf}; 2. training with references that contain the same identities \cite{liu2025manganinja,zhuang2025cobraefficientlineart}. Given the complexity of adapter-based baselines, we provide a concise overview of their operation and integration into the image-guided sketch colorization in the supplementary materials.\\



\noindent\textbf{Qualitative comparison.} We present two qualitative comparisons in Figure \ref{cross-qualitative} and Figure \ref{qualitative}. Among the baselines, MangaNinja \cite{liu2025manganinja} and Cobra \cite{zhuang2025cobraefficientlineart} are specifically designed for character colorization. These methods mitigate spatial entanglement artifacts by using training references composed of different images featuring the same identities. However, this strategy often leads to overfitting, resulting in notable performance degradation when the input sketch and reference image are semantically or geometrically misaligned. This limitation reduces the generalization ability of their models. Consequently, their models exhibit limited generalization. This issue is evident in Figure \ref{qualitative}, where results in rows (a)–(g) show significantly lower visual quality compared to those in rows (h)–(j). The GAN-based method \cite{yan-cgf} struggles to synthesize accurate colors and backgrounds for complex inputs due to limitations in its neural backbone. Compared to these baselines, our framework consistently generates high-quality colorized images with strong similarity to the reference images while effectively avoiding artifacts across a wide range of content.

Furthermore, we exhibit cross-content colorization of figure and landscape sketches in Figure \ref{cross-qualitative}. This scenario lacks clear correspondence between input identities, so subjective evaluations usually prioritize the similarity of style, color scheme, and texture/stroke details, all of which require a reasonable transfer of low-level visual features. The proposed framework significantly outperforms baseline methods in this regard while closely adhering to the sketch semantics to achieve clearer visual segmentation outcomes. A further comparison with ColorizeDiffusion \cite{cvpr-colorizediffusion} regarding style transfer is contained in the supplementary materials. \\

\noindent\textbf{Quantitative comparison.} Due to the lack of a batch inference script in Cobra official implementation \cite{zhuang2025cobraefficientlineart}, we excluded it in this comparison. Aesthetic score \cite{aesthetic-predictor} and Fréchet Inception Distance (FID) \cite{HeuselRUNH17} are widely used quantitative metrics to evaluate the perceptual quality of synthesized results without requiring inputs to be semantically and spatially paired. We conducted two evaluations using these metrics on the entire validation set, which contains 52k+ (sketch, reference) image pairs. Reference images were randomly selected during the validation. Besides, we also tested multi-scale structural similarity index measure (MS-SSIM), peak signal-to-noise ratio (PSNR), and CLIP score \cite{RadfordKHRGASAM21,openclip} for assessing the similarity between generated images and given ground truth. As these metrics require the reference to be aligned with ground truth, we selected 5000 color images as ground truth to generate extracted sketches and deformed references, where references were deformed using thin plate spline (TPS) transformation. We show the results of quantitative evaluation in Table \ref{quantitative}, where the proposed method significantly outperforms in all evaluations owing to the removal of artifacts, higher fidelity to the sketch composition, and stronger style transfer ability.\\

\begin{figure}[t]
    \centering
    \includegraphics[width=\linewidth]{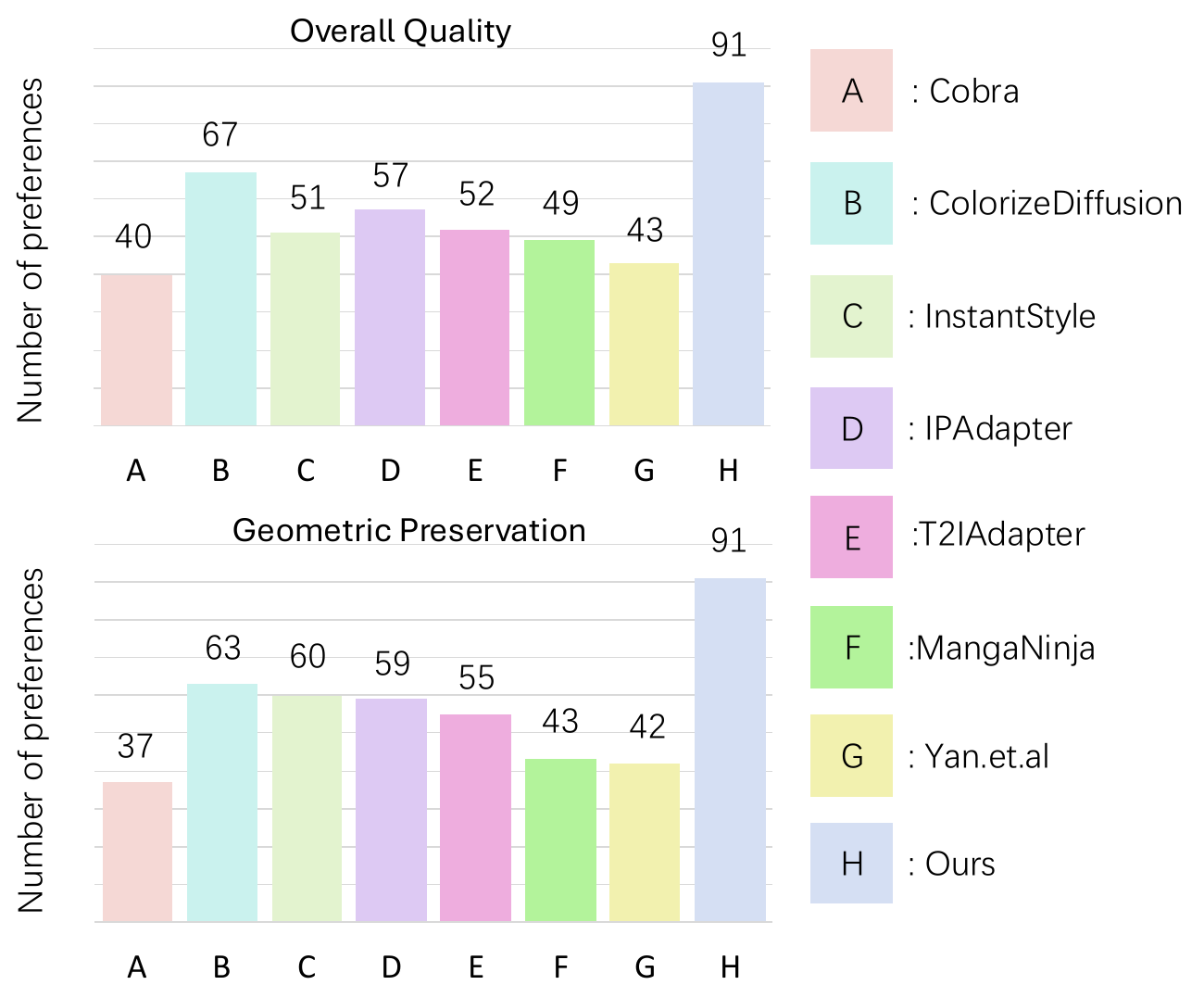}
    \vspace{-1.5em}
    \caption{Results of user study. Our method is preferred across all shown methods in overall quality and geometric preservation.}
    \vspace{-1.5em}
    \label{userstudy}
\end{figure}

\noindent\textbf{User study.} To further reveal the subjective evaluation of the proposed method and existing methods by real persons, we demonstrate a user study with 30 participants from Anime lovers communities invited to select the best results with two criteria: the overall colorization quality and the preservation of the geometric structure of the sketches. 35 image sets are prepared, and each participant is shown 15 image sets for evaluation. We present to participants the colorization results of the proposed method and those generated by six existing methods for each image set. We present the results of the user study in Figure \ref{userstudy}, with the results showing that our proposed method has received the most numbers of preferences across all the methods illustrated. For further validation of the comparison, the Kruskal-Wallis test is employed as a statistical method. The results demonstrate that our proposed method outperforms all previous methods significantly in terms of user preference with a significance level of p \textless 0.05. All the images shown in the user study are included in the supplementary materials.
\section{Conclusion}

This paper presents an image-guided sketch colorization framework designed to achieve state-of-the-art results with arbitrary inputs. To address the limitations of existing methods, we introduce the multi-stage framework and employ two independent encoders: one for background transfer and another for style modulation. This architecture enables zero-shot colorization for references in any style, even for realistic images.

The main limitations of this framework are high-precision character colorization and disentanglement of sketch-guided foregrounds, which are discussed in details in the supplementary materials. Our future work will focus on improving identity consistency and disentanglement in character colorization.

{\small
\bibliographystyle{ACM-Reference-Format} 
\bibliography{egbib}
}

\end{document}